# Levenshtein Distance Technique in Dictionary Lookup Methods: An Improved Approach

**Rishin Haldar and Debajyoti Mukhopadhyay**

Web Intelligence & Distributed Computing Research Lab
Green Tower, C-9/1, Golf Green, Calcutta 700095, India
Email: {rishinh, debajyoti.mukhopadhyay}@gmail.com

## Abstract

*Dictionary lookup methods are popular in dealing with ambiguous letters which were not recognized by Optical Character Readers. However, a robust dictionary lookup method can be complex as apriori probability calculation or a large dictionary size increases the overhead and the cost of searching. In this context, Levenshtein distance is a simple metric which can be an effective string approximation tool. After observing the effectiveness of this method, an improvement has been made to this method by grouping some similar looking alphabets and reducing the weighted difference among members of the same group. The results showed marked improvement over the traditional Levenshtein distance technique.*

## 1. Introduction

Significant research has been done in the field of Optical Character Recognition (OCR). However, there always exists a margin of error, however small and minute it is. In order to further reduce the error, many techniques are applied on the results [1][3][4][5]. Dictionary lookup methods are a simple and effective way for approximate string matching and thus are used as a supplement in the post processing phase. Dictionary lookup methods take a string as input, and tries to find the closest match(es) with entries in the dictionary. This is particularly effective, when the OCR fails to recognize the word correctly (there are some ambiguous letters), and the string can be taken as input for the dictionary lookup methods. This effort observes the effectiveness of a specific dictionary lookup technique (after some initial survey) and then look for possibilities to improve.

Within dictionary lookup methods [1], there are some issues which need to be addressed, such as:
1) Value additions to dictionary lookup have resulted in calculating apriori probabilities, which increases the overhead and complexity.
2) A short dictionary might not be sufficient to match the word in context.
3) A large dictionary increases the cost of searching to an enormous degree.

In order to avoid the overhead of calculating the apriori probabilities, and still come up with an effective dictionary lookup, we focused on approximation in string matching by using a metric called Levenshtein distance [6][7][8]. After examining the effectiveness of this method, some possible modifications are made to the algorithm, which improves the output significantly without increasing much overload.

## 2. Existing Method

### 2.1 Levenshtein Distance

Levenshtein distance (LD) is a measure of the similarity between two strings, the source string (s) and the target string (t). The distance is the number of deletions, insertions, or substitutions required to transform s into t. The greater the Levenshtein distance, the more different the strings are. In our case, the source string is the input, and the target string is one of the entries in the dictionary.

Intuitively "GUMBO" can be transformed into "GAMBOL" by substituting "A" for "U" and adding "L" (one substitution and one insertion = two changes).

## 2.2 The algorithm

Step 1: Initialization
a) Set n to be the length of s, set m to be the length of t.
b) Construct a matrix containing 0..m rows and 0..n columns.
c) Initialize the first row to 0..n,
d) Initialize the first column to 0..m.

Step2: Processing
a) Examine s (i from 1 to n).
b) Examine t (j from 1 to m).
c) If s[i] equals t[j], the cost is 0.
d) If s[i] doesn't equal t[j], the cost is 1.
e) Set cell d[i,j] of the matrix equal to the minimum of:
    i) The cell immediately above plus 1: d[i-1,j] + 1.
    ii). The cell immediately to the left plus 1: d[i,j-1] + 1.
    iii The cell diagonally above and to the left plus the cost: d[i-1,j-1] + cost.

Step 3: Result
Step 2 is repeated till the d[n,m] value is found

### 2.2.1 An Example

Finding Levenshtein Distance between GUMBO and GAMBOL [9]:

**Table 1.1.** Step 1, iteration (i) = 0

|   |   | G | U | M | B | O |
|---|---|---|---|---|---|---|
|   | 0 | 1 | 2 | 3 | 4 | 5 |
| G | 1 |   |   |   |   |   |
| A | 2 |   |   |   |   |   |
| M | 3 |   |   |   |   |   |
| B | 4 |   |   |   |   |   |
| O | 5 |   |   |   |   |   |
| L | 6 |   |   |   |   |   |

**Table 1.2.** Step 2, iteration (i) = 1

|   |   | G | U | M | B | O |
|---|---|---|---|---|---|---|
|   | 0 | 1 | 2 | 3 | 4 | 5 |
| G | 1 | 0 |   |   |   |   |
| A | 2 | 1 |   |   |   |   |
| M | 3 | 2 |   |   |   |   |
| B | 4 | 3 |   |   |   |   |
| O | 5 | 4 |   |   |   |   |
| L | 6 | 5 |   |   |   |   |

**Table 1.3.** Step 3, iteration (i) = 2

|   |   | G | U | M | B | O |
|---|---|---|---|---|---|---|
|   | 0 | 1 | 2 | 3 | 4 | 5 |
| G | 1 | 0 | 1 |   |   |   |
| A | 2 | 1 | 1 |   |   |   |
| M | 3 | 2 | 2 |   |   |   |
| B | 4 | 3 | 3 |   |   |   |
| O | 5 | 4 | 4 |   |   |   |
| L | 6 | 5 | 5 |   |   |   |

Finally,

**Table 1.** Last step, iteration (i) = n, j=m

|   |   | G | U | M | B | O |
|---|---|---|---|---|---|---|
|   | 0 | 1 | 2 | 3 | 4 | 5 |
| G | 1 | 0 | 1 | 2 | 3 | 4 |
| A | 2 | 1 | 1 | 2 | 3 | 4 |
| M | 3 | 2 | 2 | 1 | 2 | 3 |
| B | 4 | 3 | 3 | 2 | 1 | 2 |
| O | 5 | 4 | 4 | 3 | 2 | 1 |
| L | 6 | 5 | 5 | 4 | 3 | 2 |

The distance is in the lower right hand corner of the matrix, i.e., 2.

# 3. Modifications

As pointed out by the algorithm, in Levenshtein distance, the difference between different literals are uniform (i.e., 1). However, if certain similar shaped literals can be identified and given different weight difference (< 1), then nearest matches will be more accurate. e.g., O, D, Q can be given a weight of 0.4 instead of 1. Thus, the code was modified for Levenshtein distance to incorporate these new weights and group information.

The groups identified are
1) O, D, Q
2) I, J, L, T
3) U, V
4) F, P
5) C, G

## 3.1 Justification of this modification

This bias utilizes the nature of human handwriting. For example, if the word that the user wanted to write is BODY, but because of unclear handwriting or inefficient OCR, the result came out to be BDQY. Figure 1 explains the situation clearly. Levenshtein distance is going to give many possible words / answers. From BDQY all of the four words (BODY, BUSY, BURY, BONY) are of distance two. However, using this modified scheme, since D, Q are in the same group as O, D the distance between BDQY and BODY comes out to be the shortest compared to the others, therefore BODY is chosen as the answer. This happens to be the same word that the user attempted to write in the first place.

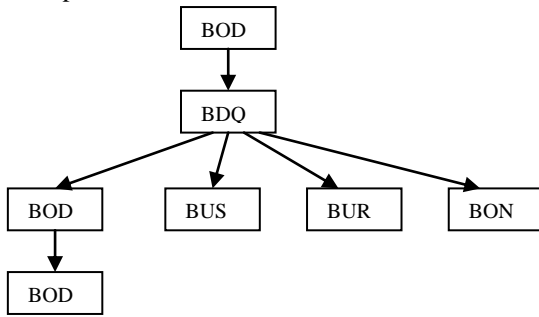

**Figure 1.** An example of possible outcomes. When the OCR outputs BDQY, normal Levenshtein distance method gives four possible outcomes, while the modified approach gives the right answer.

**Note** : Incidentally, this bias turned out to be helpful for the example in Figure 1. There is a possibility that this bias might not be of additional help (i.e., even with modified weightage, the difference from the source is identical for more than one outputs). When the normal Levenshtein distance method is used, the algorithm is tailored to output three/four closest matches, thus not restricting the answer to one single match. The modified Levenshtein distance method does not make any change in this regard, so by the modified approach also, the same output is generated. Thus if the correct answer is present in one of the outputs of Levenshtein distance method, it would also be present in the output of this modified approach.

# 4. Experimental Setup

## 4.1 Data collection

The source of the training data was the handwritten numerical data available along with the OCR system in Statistical Pattern Recognition Toolbox (STPRTools) in Matlab. There are publicly available alphabet data set from NIST and USPS resource, however, the obvious hurdles were that many of NIST databases now charge money and also pixel files from these databases were in different formats or different matrix configurations. The OCR system in STPRTools, on the other hand, consistently uses 16x16 pixels for each alphabet and the GUI provided by the OCR system in STPRTools allows the user to manually enter training samples conveniently. Therefore, this seemed like a better option.

In order to populate the dictionary, some words were acquired from /usr/dict/words on lisa.cs.mcgill.ca. This has a repository of more than 25,000 English words.

## 4.2 Data Preprocessing

As the input training samples were entered manually, there is no legitimate issue of data preprocessing. The only part of data cleaning was the acceptance of English words of length 3 or 5. Any other length word or non-word literal was not stored as a training sample. The length

(three or five) was chosen just for simplification in dictionary search.

Secondly, the numbers of training samples were reduced from 50 to 20, to reduce the effectiveness of the provided SVM classifier functions, and thus give more opportunity to the Levenshtein distance methods to observe the post processing effectiveness.

### 4.3 Software Used

Matlab's STPRTools toolbox was used, and code for both Levenshtein distance method and modified Levenshtein distance method was incorporated in Matlab.

### 4.3.1 Steps followed

1. The GUI provided by the OCR system in STPRTools was used to provide test alphabets. Entered 250 English words of length three and 250 words of length five.

2. The SVM functions available in the OCR system in STPRTools was utilized to initially recognize the handwritten pixels.

3. For those words which were not properly recognized by SVM, the Levenshtein distance method was used to reduce the number of unrecognized words.

4. For the same set of words, not recognized by SVM, the modified Levenshtein distance method was used to reduce the number of unrecognized words.

5. The traditional dictionary lookup method uses Bayesian probability to find out the three nearest words [2]. This procedure was also used to compare its effectiveness with SVM + Levenshtein Distance method.

## 5. Experimental Results

1. The provided OCR was not able to correctly recognize 93 out of 500 test inputs after processing the data through SVM.

2. The standard Levenshtein distance (LD) method, when applied to the unrecognized words, reduced the number to 66.

3. The modified method (MLD) reduced the number to 52 from 93.

The results are as follows:

**Table 2**. Unrecognized words with LD and MLD after processing by SVM

Word length of three

| Word Length 3 | Total 250 words | |
|---|---|---|
| Unrecognized | After LD | After MLD |
| 35 | 23 | 18 |

Word length of five

| Word Length 5 | Total 250 words | |
|---|---|---|
| Unrecognized | After LD | After MLD |
| 58 | 43 | 34 |

The same LD method and MLD methods were applied to unrecognized words after applying the Bayesian method [2]. Out of the total 500 words, The LD method could recognize 263 words (500 − (110+127)), whereas the modified method, MLD, applied in a similar fashion, recognized 304 words (500 − (84+112)).

**Table 3**. Unrecognized words with LD and MLD after processing by Bayesian Method

Word length of three

| Word Length 3 | Total 250 words | |
|---|---|---|
| Unrecognized | After LD | After MLD |
| 132 | 110 | 84 |

Word length of five

| Word Length 5 | Total 250 words | |
|---|---|---|
| Unrecognized | After LD | After MLD |
| 143 | 127 | 112 |

## 6. Conclusion

It is obvious from the result that this modification reduces the error significantly. Based on this approach, we believe that this significant reduction would be valid for larger number of test samples also. We have also observed that even if this approach does not narrow down the possible outcomes, the results would always be as good as the normal LD method, but never worse.

We have applied this approach to words of length three and five only, its effectiveness on words of varying/more lengths are yet to be verified. As the number of words in the dictionary grows, we have to keep in mind the optimal balance between the size of the dictionary and the overhead needed to compute, in order to make these dictionary lookup methods more effective.

From the semantic web viewpoint, while uploading / downloading scanned manuscripts, this facility can be used as a web service. Moreover, nowadays many websites generate a random sequence of alphanumeric characters which needs to be entered by the user as a means of authentication. Many a times, those characters are unclear, in those cases this approach can make the process simpler for the user.